\documentstyle[12pt,titlepage]{article}

\font\grande=cmr10 scaled \magstep4
\font\medio=cmr10 scaled \magstep2
\outer\def\beginsection#1\par{\medbreak\bigskip
      \message{#1}\leftline{\bf#1}\nobreak\medskip
\vskip-\parskip
      \noindent}

\def\laq{\raise 0.4ex\hbox{$<$}\kern -0.8em\lower 0.62
ex\hbox{$\sim$}}
\def\gaq{\raise 0.4ex\hbox{$>$}\kern -0.7em\lower 0.62
ex\hbox{$\sim$}}

\begin{document}
\bibliographystyle {unsrt}
\newcommand{\pa}{\partial}

\titlepage
\begin{flushright}
CERN-TH/95-102 \\
DFTT-28/95
\end{flushright}
\vspace{15mm}
\begin{center}
{\grande Electromagnetic Origin of the CMB}\\
\vspace{5mm}
{\grande Anisotropy in String Cosmology}

\vspace{10mm}

M. Gasperini \\
{\em Dipartimento di Fisica Teorica, Via P. Giuria 1, 10125 Turin,
Italy} \\
M. Giovannini and G. Veneziano \\
{\em Theory Division, CERN, CH-1211 Geneva 23, Switzerland} \\
\end{center}
\vspace{10mm}
\centerline{\medio  Abstract}

\noindent
In the inflationary scenarios suggested by string theory, the
vacuum fluctuations of the electromagnetic field can be amplified by
the time-evolution of the dilaton background, and
can grow large
enough to explain both the origin of the cosmic magnetic fields and
of the observed CMB anisotropy. The normalization of the perturbation
spectrum is fixed, and implies a relation between the perturbation
amplitude at the COBE scale and the spectral index
$n$. Working within
a generic two-parameter family of backgrounds, a large scale anisotropy
$\Delta T /T \simeq 10^{-5}$ is found to correspond to
a spectral index in the range
$n\simeq 1.11 - 1.17$.

\vspace{5mm}
\vfill
\begin{flushleft}
CERN-TH/95-102 \\
April 1995 \end{flushleft}

\newpage

\setcounter{equation}{0}

In the standard inflationary scenario, the anisotropy of the Cosmic
Microwave Background (CMB) recently detected by COBE
\cite{Smoot,Bennet}
is usually attributed to the cosmological amplification of the quantum
fluctuations of the metric. These consist of both tensor
(gravitational waves) and scalar perturbations, the latter being
coupled to the energy density fluctuations.
 The observed inhomogeneities of the CMB radiation
could also emerge from the vacuum fluctuations of the
electromagnetic radiation itself, through their contribution to
${\delta \rho}/{\rho}$. However, the minimal coupling of
photons to the metric is conformally invariant in $d=3$ spatial
dimensions, and it is difficult, in general, to obtain a significant
amplification of the electromagnetic fluctuations in the context of
the standard inflationary
scenario \cite{Turn}.

In the inflationary models based on the low energy limit of critical
superstring theory \cite{Ven,Astr,GV}, the electromagnetic field
$F_{\mu\nu}$ is coupled not only to the metric ($g_{\mu\nu}$) but
also to
the dilaton ($\phi$) background, according to the dimensionally
reduced,
effective action \cite{Lov}
\begin{equation}
S=- \int d^4x |det(g_{\mu\nu})|^{1/2}
e^{-\phi}( R +
\partial_{\mu} \phi \partial^{\mu} \phi
+ \frac{1}{4} F_{\mu\nu}F^{\mu\nu})
\label{1}
\end{equation}
In this context $\phi$, which controls the tree-level
four-dimensional
gauge coupling $g^2=e^\phi$, is rapidly changing in time  and can
amplify
directly the electromagnetic vacuum fluctuations \cite{Mag,Lem}.
On the other hand, tensor metric perturbations, as well as the scalar
perturbations induced by dilatonic fluctuations, are characterized in
this context by
``blue" spectra strongly tilted towards large frequencies \cite
{GV,MG,Brus}, with an amplitude on large angular scales which is far
too
low to match  COBE's observations. It becomes thus crucial, for the
purpose of testing string theory through its astrophysical
consequences
\cite{Proc}, to decide whether or not the CMB anisotropy could
originate
from the vacuum quantum fluctuations of the electromagnetic field
itself,
after they have been amplified by the time-evolution of the dilaton
background.

The purpose of this paper is to show that, in an appropriate range
 for two arbitrary parameters characterizing
 a generic string cosmology scenario, such an electromagnetic origin
of
the anisotropy is possible, is consistent with the linearization of
perturbations around a nearly homogeneous background, and is also
consistent with various phenomenological constraints (following from
pulsar timing data and nucleosynthesis). In that range of parameters,
moreover, the same mechanism that amplifies the electromagnetic
vacuum
fluctuations can also be responsible for the production of the
observed
galactic (and extragalactic) magnetic fields \cite{Mag}, and can thus
explain why the average energy density of the cosmic magnetic fields
and
of the CMB radiation are of the same order.

Let us consider the evolution of
the quantum fluctuations of the electromagnetic field, according to
the
action (\ref{1}). In a four-dimensional,
conformally flat background,
the  Fourier modes $A_{k}^{\mu}$ of the (canonically
normalized) electromagnetic variable
 satisfy the equation
\begin{equation}
A_{k}^{\prime\prime}+
[k^2-V(\eta)]A_{k}=0~~~~,~~~~V(\eta)=
g(g^{-1})^{\prime\prime}~~~~,~~~~
g(\eta)\equiv e^{\phi /2}
\label{2}
\end{equation}
This equation is valid for each polarization component, and
 is obtained from the action (\ref{1}) with the gauge
condition
$\partial_{\nu}[e^{-\phi}\partial^{\mu}(e^{\frac{\phi}{2}}
A^{\nu})]=0$
(a prime denotes
differentiation with respect to the conformal time $\eta$). Note
 the analogy with the tensor part of the
metric perturbation equations \cite{Gris}, which has the same form as
(\ref{2})
with the inverse of the coupling, $g^{-1}$, replaced simply
by the Einstein-frame scale factor $a_{E}=g^{-1} a$.

In our context
$V(\eta)$ represents an effective potential barrier, approaching zero
as $\eta \rightarrow \pm \infty$. A mode of comoving frequency $k$,
``hitting" the barrier at the time $\eta=\eta_{ex}(k)$, is thus
parametrically
amplified just like in the case of tensor perturbations. The modulus
of
the Bogoliubov coefficient $|c(k)|$ describing this amplification
turns out to be given, to leading order, by the ratio of the gauge
coupling at reentry
and at exit \cite{Mag}
\begin{equation}
|c(k)|\simeq\frac{g_{re}}{g_{ex}}\equiv
\exp\{-{1\over 2}[\phi(\eta_{ex})-\phi(\eta_{re})]\}
\label{3}
\end{equation}
where $\eta_{ex}(k)$ and $\eta_{re}(k)$
are defined by
$k^2 = |V(\eta_{ex})| = |V(\eta_{re})|$.
The Bogoliubov coefficient $c(\omega)$
defines the energy distribution $\rho(\omega)$ of
the amplified fluctuation spectrum, through the relation
$ d\rho/d\ln\omega \simeq ({\omega^4}/{16 \pi^2})
|c(\omega)|^2$, where $\omega (t)= k/a(t)$ is the red-shifted,
present value of the
amplified proper frequency. We are
interested, in particular, in the ratio
\begin{equation}
r(\omega)=\frac{\omega}{\rho_{cmb}} \frac{d\rho}
{d\omega}
\simeq
\frac{\omega ^{4}}{16 {\pi}^2 \rho_{cmb}}
\left[g_{re}(\omega)\over g_{ex}(\omega)\right]^2
\label{4}
\end{equation}
measuring the fraction of electromagnetic energy stored in the
mode $\omega$,
 relative to the CMB energy density $\rho_{cmb}$.

In order to compute this ratio, we must use the explicit time
evolution of the dilaton background, as predicted by the inflationary
models based on the string effective action \cite{Astr,GV}. In such
models the dilaton
undergoes an accelerated evolution
from the string perturbative vacuum ($g=0$, $\phi= -\infty$) towards
the
strong
coupling regime, where it is expected to remain frozen at its
present value ($g=g_1=e^{\phi_1/2}=$ const). The initial phase
of growing  curvature and dilaton coupling (also called
``pre-big-bang"
scenario \cite{Astr,GV}) is
driven by the kinetic energy of the
dilaton field (with negligible contributions from the dilaton
potential), and can be
described in terms of
 the lowest order string effective action only
up to the time $\eta=\eta_{s}$  at which the curvature
reaches the string scale $H_{s}=\lambda_{s}^{-1}\equiv
(\alpha^{\prime})^{-1/2}$ [determined by the string tension
$(\alpha^{\prime})^{-1}$]. The value $\phi_{s}$ of the dilaton
at $\eta=\eta_s$ is the first important parameter of our scenario.
Provided such
 value is sufficiently negative
it is also arbitrary, since we are still in the perturbative regime
at
$\eta=\eta_s$, and
there
is no perturbative potential to break invariance under shifts of
$\phi$.

For $\eta >\eta_{s}$, however, higher
orders in $\alpha^{\prime}$
become important in the string effective action,
and the background enters a genuinely ``stringy" phase of
 unknown duration, assumed to end at $\eta=\eta_1$ with a smooth
transition to the standard radiation-dominated regime (where
$\phi=\phi_1=$const). As shown in \cite{BV},
it is impossible to have a graceful
exit to  standard cosmology without such an intermediate
stringy phase, after which the dilaton,
feeling a non-trivial potential, is attracted to its
present constant value.
The total red-shift $z_s=a_1/a_s$
encountered during
 the stringy epoch,
will be the second crucial parameter (besides $\phi_{s}$)
entering  our discussion. For our purpose,
two
parameters are enough to specify completely our model of
background, if we accept that during the string phase the
curvature stays controlled by the string scale, that is $H\simeq
\lambda_{s}^{-1}$ for $\eta_s<\eta <\eta_1$.

We will work in the so-called String frame, in which
the string scale $\lambda_{s}$
 is constant, while the Planck scale
$\lambda_{p}= \lambda_{s} e^{\phi/2}$ grows from zero
(at the initial vacuum) to its present value $\lambda_p\simeq
10^{-19}
(GeV)^{-1}$  reached at the end of the
string phase. In the low energy phase driven by the dilaton, the
dilaton
evolution is exactly known \cite{Ven,Astr,GV} and is given, in the
String frame, by
\begin{equation}
\phi=(3 + \sqrt{3})\ln a + const = -\sqrt{3} \ln |\eta| + const ,
{}~~~~~~~~~~~~~~~~~\eta<\eta_s
\label{5}
\end{equation}
Internal dimensions affects slightly the above numerical constants
without affecting the results of this paper. During the string phase
the curvature stays constant ($H\simeq
\lambda_s^{-1}$) so that, in this frame, the string epoch is
characterized by a de Sitter-like evolution of the metric background,
with $\eta_s/\eta_1=a_1/a_s\equiv z_s$. In addition, the  rate of
growth of
the coupling during the string phase should also be bounded
by the string scale (like the space-time curvature). Defining
$\dot\phi
\simeq 2\beta H\simeq 2\beta \lambda_s^{-1}$, where $\beta$ is some
constant of order of unity [for instance, eq.(5) gives $\beta
\simeq 2.37$ in the
dilaton driven epoch], the ``average" time behaviour of the dilaton
between $\eta_s$ and $\eta_1$ can be parameterized as
\begin{equation}
\phi= -2\beta \ln |\eta| + const ,
{}~~~~~~~~~\beta= -{(\phi_s-\phi_1)\over 2 \ln z_s} ,
{}~~~~~~~~~\eta_s < \eta < \eta_1
\label{6}
\end{equation}

For this model of background, the effective potential $V(\eta)$ of
eq.(\ref{2}) grows like $\eta^{-2}$ for $\eta \rightarrow 0_-$ in the
dilaton-driven phase, keeps growing during the string phase , where
it
reaches a maximal value $\sim \eta_1^{-2}$ around the final time
$\eta_1$, and then goes rapidly to zero at the beginning of the
radiation dominated era, where $g(\eta)=g_1=$const. A mode
hitting the barrier (or ``crossing
the horizon") during the dilaton phase thus remains under the barrier
during the whole string phase. As a consequence, $\eta_{re}>\eta_1$
and
$\phi_{re}(\omega)=\phi_1=$const for all $\omega$.

The spectral distribution $r(\omega)$ is now completely fixed in
terms
of our two parameters $\phi_s$ and $z_s$. Using
$\rho_{cmb}(\eta_{1})\simeq M_{p}^2 H_{1}^2 \simeq g_{1}^{-2}
H_{1}^4$, ($M_p$ being the present value of the Plank Mass),
eq.(\ref{4}) leads simply to
\begin{equation}
r(\omega)\simeq \frac{g_1^2}{16 \pi^2} \left(\omega \over \omega_1
\right)^4
e^{-\Delta \phi_{ex}(\omega)}
\label{7}
\end{equation}
where $\Delta \phi_{ex}(\omega)=\phi_{ex}(\omega)-\phi_1$, and
$\omega_1
=H_1a_1/a\simeq (g_1/4 \pi)^{1/2}10^{11}$Hz is
the maximal amplified
frequency
(the amplitude of modes $\omega > \omega_1$ is exponentially
suppressed,
and will be neglected throughout this paper). For modes
$\omega>\omega_s
\equiv\omega_1/z_s$, crossing the horizon during the string phase, we
thus obtain the spectrum
 \begin{equation}
r(\omega)\simeq  \frac{g_{1}^2}{16 \pi^2}
\left(\frac{\omega}{\omega_{1}}\right)
^{4 -2\beta}
{}~~~~~~,~~~\omega_{s}<\omega<\omega_{1}
\label{8}
\end{equation}
For modes crossing the horizon in the dilaton phase ($\omega<
\omega_s$) we have instead, from eqs.(\ref{5},\ref{6},\ref{7}),
 \begin{equation}
r(\omega)\simeq \frac{g_{1}^2}{16 \pi^2} \left(
\frac{\omega}{\omega_{1}}\right)^{4-\sqrt 3}
z_{s}^{-\sqrt 3} e^{-\phi_{s}}
{}~~~~\,\,\,\,\,\,\,,~~~\omega<\omega_{s}
\label{9}
\end{equation}

The above electromagnetic spectrum has been obtained by using a
homogeneous and isotropic model of metric (and dilaton) background.
It
is thus valid provided the amplified fluctuations remain, at all
times, small perturbations of a nearly homogeneous configuration, with
a negligible back-reaction on the metric. This requires $r
(\omega)\laq 1$, at all $\omega$.
This bound is satisfied, for $(g_1/4\pi) \laq 1$, provided
\begin{equation}
(g_s/g_1) \gaq z_s^{-2}
\label{10}
\end{equation}
which restricts the allowed region in the two-dimensional parameter
space ($z_s,g_s$) of our background model.

Consider now a length scale $\omega^{-1}$ reentering the horizon in
the
radiation era. The associated electromagnetic perturbation
represents,
at the time of reentry, a coherent field over the horizon scale
which,
consistently with the bound (\ref{10}), could be strong enough to
seed
the galactic dynamo mechanism, or the galactic magnetic field itself
\cite{Mag}. Soon after reentry, however, the perturbation may be
expected to thermalize and homogenize rapidly since, unlike metric
(scalar and tensor) perturbations, photons are not  decoupled from
matter in the radiation era, and the shape of their spectrum remains
frozen only outside the causal horizon.

For all scales reentering the horizon after the decoupling epoch,
however,
 the electromagnetic perturbations can contribute to the
inhomogeneity of the CMB radiation, with a spectral distribution
$\rho=r(\omega)$ determined by eqs.(\ref{8},\ref{9}).
Such a spectrum grows
with frequency, with the position of its peak fixed in the plane
($\omega, r$) (i.e. $r\simeq {g_{1}^2}/{16 \pi^2}$ at
$\omega\simeq \omega_{1}$). The
perturbation amplitude $r(\omega)$ at a given scale $\omega$ can thus
be
uniquely determined as a function of the unknown duration and slope
of
the ``stringy" branch (\ref{8}) of the spectrum, namely in terms of
the
two parameters $z_s$, $g_s$.

A  large enough perturbation to match COBE's observations \cite
{Smoot,Bennet}, $\Delta T/T\simeq 10^{-5}$ at the present
horizon scale $\omega_0$ would require a spectral energy density such
that, in critical units, $\Omega(\omega_0)\equiv
\rho_c^{-1}[d\rho(\omega)/d\ln
\omega]_{\omega=\omega_0}
\simeq 10^{-10}$. In terms of our variable $r(\omega)$ this
implies
\begin{equation}
r(\omega_0)\simeq 10^{-6} , ~~~~~~~~~~~~~~~~~~~~~\omega_0
\simeq 10^{-18} Hz
\label{11}
\end{equation}
If the scale $\omega_0$ crossed the horizon during the dilaton phase
(i.e. if $z_s\laq 10^{29}$),
this condition is compatible with eq.(\ref{10}) only
in a very small region of parameter space. In such case an
electromagnetic origin of the CMB anisotropy is possible, but
requires
 a certain degree of fine-tuning. If, on the contrary, the horizon
crossing of $\omega_0$
occurred during the string phase ($z_s \gaq 10^{29}$), and we define as
usual the
spectral index $n$ for eq.(\ref{8}) as $n-1=4-2\beta$, then the
present electromagnetic contribution to the
anisotropy at the scale $\omega_0$ can be
written as
\begin{equation}
\log_{10} r(\omega_0)\simeq -29(n-1) + {1\over 2}(5-n)
\log_{10}(\frac{g_{1}}{4 \pi}) ,~~~~~~~~~~~
z_s\gaq 10^{29}
\label{12}
\end{equation}
This equation (which is the main result of this paper) relates the
perturbation amplitude at the scale $\omega_0$ to the spectral
index $n$, and provides a
condition on the parameter space ($z_s,\phi_s$) which is always
compatible with eq.(\ref{10}), as $(g_1 /4 \pi) \leq 1$.
The requirement (\ref{11}), in particular, is satisfied for
\begin{equation}
n \simeq {35+{5\over 2}\log_{10}(g_1 /4\pi) \over 29+
{1\over 2}\log_{10}(g_1 /4\pi)}
\label{13}
\end{equation}

Typical values of $(g_1 /4\pi)^2$ range from $10^{-1}$
to $10^{-3}$ \cite{gabriele}. The COBE observations are thus accounted
for, in this context, for values of the spectral index that are
typically in the range
$n\simeq 1.11-1.17$.
Such a spectral index is certainly flat enough to be well consistent
with
the analysis of the first two years of the COBE DMR data
\cite{Bennet}.
It may be worth recalling that slightly growing ($n>1$, also called
``blue") spectra, like this,
have been invoked \cite{Piran}
to explain the claimed bulk flow and large voids in the galaxy
distribution, on scales of order $10^2$Mpc (see also \cite{Moll}).
Our range of
values, moreover, is consistent with the upper bound
$n<1.5$ recently obtained by using the COBE FIRAS limits on the CMB
spectral distortions \cite{Hu}. Note that, according to
eq.(\ref{13}), $n$ depends very weakly on the precise value of $g_1$, so
that our estimate is quite stable, in spite of the rather large
theoretical uncertainties about $g_1$. Note also that the above value of
$n$ corresponds to an average value of $\dot \phi/2H\simeq \beta$ of
about $1.9$, which is not far from the value $2.37$
characteristic of the
dilaton-driven era.

In order to give some concrete estimate of the phenomenological
bounds
relevant for the problem
we will set $g_{1}/4 \pi \simeq 1$ in the following [if the value of
$(g_{1}/4\pi)^2$ would be
smaller the constraints
which we will discuss will be satisfied even better,  thanks to
the
flatter ensuing  spectrum].
In such case, the CMB anisotropy can receive
a complete
electromagnetic explanation
provided the parameters of our background are constrained to lie on
(or near) the half-line $\log_{10} g_s =-1.90\log_{10}z_s$,
$\log_{10}z_s>29$. The resulting perturbation spectrum
\begin{equation}
r(\omega)\simeq \left( \omega \over 10^{11}Hz \right)^{6/29} ,
{}~~~~~~~~~~~~~~~\omega <10^{11}Hz
\label{14}
\end{equation}
gives then $r\simeq 10^{-5}$ at the intergalactic scale $\omega_G
=(1 Mpc)^{-1}$, which is large enough to seed not only the galactic
dynamo, but also the cosmic magnetic field directly \cite{Mag}. This
results holds in general for any realistic value of $g_1$, and
leads to consider a scenario in which
the CMB anisotropy and the primordial magnetic fields have a
common origin. The peak value of order unity  of the electromagnetic
spectrum can then easily explain the (otherwise mysterious, to the best
of our
knowledge) coincidence that the total energy density of our galactic
magnetic field, $\rho_B =\rho_{cmb} \int^{\omega_1}
r(\omega)d\omega$,
is of the same order of magnitude as the CMB energy density.
We also note that, at the scale corresponding to the end of
nucleosynthesis ($\omega_N\simeq 10^{-12}$Hz), eq.(\ref{14}) predicts
$r\simeq 10^{-4.7}$. It is thus automatically consistent with the
bounds
following from the presence of strong magnetic fields at
nucleosynthesis
time \cite{Rub}, which impose $r(\omega_N)\laq 0.05$.

The above discussion refers to the case in which all scales inside
our
present horizon crossed the horizon, for the first time, during the
string phase, i.e. for $z_s\geq 10^{29}$. Such a phase was
characterized by
a de Sitter-like metric evolution, with a curvature scale of
Planckian
order. In the standard inflationary context such a background
configuration is forbidden, as it would lead to an overproduction of
tensor perturbations: a phase of constant Planckian curvature can
last
only up to a total red-shift $z\leq 10^{19}$ \cite{MG2}, to be
compatible with
the bounds obtained from pulsar timing data \cite{Stin}. In a
string cosmology context, however, we must recall that the de
Sitter-like evolution of the metric refers to the String frame, where
tensor metric perturbations are also coupled to the dilaton
background
\cite{MG}. As a consequence, the spectral distribution $r_g(\omega)$
of
tensor perturbation is growing with frequency (instead of being flat
like in the standard de Sitter scenario), with a peak value which is
again of order one around $10^{11}$Hz. Moreover, the growth is so
fast
($r_g\sim \omega^{n+1}$, where $n$ is the spectral index of the
electromagnetic perturbations), that the contribution of $r_g$ is
negligible at the scale $\omega_0$, and it is also largely consistent
with the pulsar bound \cite{Stin} which requires $r_g(\omega_P)\laq
10^{-2}$ at $\omega_P\simeq 10^{-8}$Hz.

A further remark related to the long duration of the string phase
concerns the validity of the spectrum (\ref{8}), which has been
obtained
from the tree-level, low energy action (\ref{1}). It is true that,
in the string phase, we may have corrections coming both from
higher loops
(expansion in $e^{\phi}$) and from higher derivative terms
($\alpha^{\prime}$
corrections ). However, in order to reproduce the large scale
anisotropy
we have to
work in a range of parameters where the dilaton is
deeply  in his perturbative
regime. Eq.(\ref{14}) holds in fact for $g_s=e^{\phi_s/2} \laq
10^{-55}$. We thus
expect our results to be stable against  loop
corrections, at least at all the scales which are relevant for the
observed anisotropy and for the generation of primordial magnetic
fields.

As to the $\alpha^{\prime}$ corrections, they are instead crucial in
the basic assumption that the dilaton driven era leads to a quasi de
Sitter epoch when
the curvature reaches the string scale. Concerning the higher
derivative corrections of the form
$(\alpha' F_{\mu\nu}F^{\mu\nu})^m$, $m\geq 2$, they can modify in
principle
the equation determining the evolution of electromagnetic
fluctuations
(eq.(\ref{2})). We are expanding, however, our perturbations around
the
vacuum background $F_{\mu\nu}=0$. Therefore, no higher curvature
correction may provide significant contributions as long as we work in
the
region of parameter space in which perturbations can be consistently
treated linearly, namely in the region in which eq.(\ref{10}) is
satisfied.

We thus believe that the main conclusion of this paper, namely that
an
electromagnetic origin of the CMB anisotropy is allowed in a
realistic
string cosmology scenario, is not only compatible with the various
phenomenological bounds, but is also  quite independent of the
(unknown) kinematic
details of the high energy ``stringy" phase, preceding the phase of
standard cosmological evolution.

\end{document}